\documentclass{JINST}

\title{Towards a direct transition energy measurement of the lowest nuclear 
       excitation in $^{229}$Th}

\author{L. v.d. Wense$^{a,b}$, 
        P.G. Thirolf$^a$\thanks{Corresponding author.}, D. Kalb$^a$ and M. Laatiaoui$^{c,d}$ \\
\llap{$^a$}Ludwig-Maximilians-Universit\"at M\"unchen,
           Am Coulombwall 1, Garching, Germany\\
\llap{$^b$}Max-Planck-Institute f. Quantenoptik, 
           Hans-Kopfermann-Str. 1, Garching, Germany,\\
\llap{$^c$}GSI Helmholtzzentrum f\"ur Schwerionenforschung GmbH, 
           Planckstr. 1, Darmstadt, Germany,\\
\llap{$^d$}Helmholtz Institut Mainz, Johann-Joachim-Becherweg 36, Germany \\

  E-mail: \email{Peter.Thirolf@physik.uni-muenchen.de}}

\abstract{The isomeric first excited state of the isotope $^{229}$Th 
exhibits the lowest nuclear excitation energy in the whole landscape of 
known atomic nuclei. For a long time this energy was reported in the literature 
as 3.5(5) eV, however, a new experiment corrected this energy to 
7.6(5) eV, corresponding to a UV transition wavelength of 163(11) nm. 
The expected isomeric lifetime is $\tau=$ 3-5 hours, leading to an extremely 
sharp relative linewidth of $\Delta E/E\approx 10^{-20}$, 5-6 orders of magnitude 
smaller than typical atomic relative linewidths. For an adequately chosen electronic 
state the frequency of the nuclear ground-state transition will be independent from
influences of external fields in the framework of the linear Zeeman and
quadratic Stark effect, rendering $^{229m}$Th a candidate for a reference of an
optical clock with very high accuracy~\cite{pei03}.
Moreover, in the literature speculations about a potentially enhanced 
sensitivity of the ground-state transition of $^{229m}$Th for eventual time-dependent
variations of fundamental constants (e.g. fine structure constant $\alpha$)
can be found~\cite{fla06,ber09}.\\
We report on our experimental activities that aim at a direct identification of 
the UV fluorescence of the ground-state transition energy of $^{229m}$Th. 
A further goal is to improve the accuracy of the
ground-state transition energy as a prerequisite for a laser-based optical
control of this nuclear excited state, allowing to build a bridge between 
atomic and nuclear physics and open new perspectives for metrological as well
as fundamental studies.}


\begin{document}

jinst-229mth-final.dvi keepcomments

\section{Introduction}

Amongst the presently about 2800 known isotopes, $^{229}$Th occupies a special
position due to its first excited state which exhibits the lowest excitation
energy that so far has been identified in any of the known atomic nuclei
(the next comparably low excitation energy can be found in $^{235}$U with 73 eV).
Since more than 30 years experiments have been performed to measure the
energy splitting of the ground-state doublet in $^{229}$Th
(with spin and parity J$^{\pi} = 5/2^+, 3/2^+$, respectively and 
Nilsson quantum numbers [633] for the
ground state and [631] for the first excited state)~\cite{kro76,rei90,hel94}, 
without so far achieving a direct observation of the $\gamma$ decay.
The so far available data all originate from an analysis of combinations
between higher-lying states that decay into the ground and first excited state,
respectively.\\
For a long time 3.5 eV was generally adopted as value of the transition energy
(in Ref.~\cite{hel94} quoted as (3.5$\pm$1.0) eV), until
 an improved experiment of the Livermore group using a novel X-ray spectrometer
resulted in a new and significantly altered value of
7.6(5) eV~\cite{bec07}, recently updated to 7.8(5) eV~\cite{bec10}.
This new value places the isomeric $^{229}$Th ground-state transition
in the spectral region of the deep UV (VUV: Vacuum UV), while the previous
experiments had concentrated on the visual range of the electromagnetic spectrum.
So far no unambiguous direct identification of this transition has been achieved.
\begin{figure}[htb]
  \centerline{\includegraphics[width=0.3\textwidth,angle=-90]
             {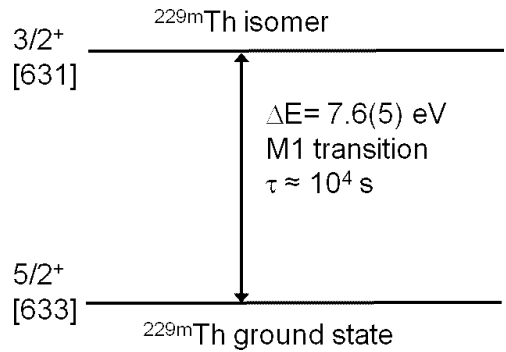}}
         \caption[]
                {Almost degenerate ground-state doublet in $^{229}$Th with the
                 energetically lowest nuclear ground state and the isomeric 
                 first excited state at 7.6(5) eV.}
        \label{fig:th229m-levels}
\end{figure}
In contrast to the ground state of $^{229}$Th with a half life of 7880 years,
the first excited state represents a long-lived isomeric state with an (expected)
half-life of some (2-4) hours. In view of the extremely low excitation energy and
the long lifetime of the $^{229m}$Th isomer, its unique properties are obvious: 
a natural linewidth of ca. 10~$\mu$Hz corresponds to an extremely sharp ground-state
transition with $\Delta E/E\approx$10$^{-20}$.\\
In general, nuclear transitions profit from the about 5 orders of magnitude 
smaller nuclear radii compared to the extension of atomic systems,
leading to a much improved shielding against external perturbing fields and 
resulting in small nuclear multipole moments.
Hereby the nuclear transition frequency is independent of all influences that
lead to level shifts solely depending on electronic quantum numbers, since those
do not change during the nuclear transition and therefore will be equally affected
in both participating nuclear states. This holds for the scalar part of the
quadratic Stark effect as the typically dominant reason of shifts of electronic
transition frequencies under the influence of static electrical fields or 
electromagnetic radiation. The remaining influence of the so-called hyperfine 
Stark-drift can be estimated in the optical frequency range according to 
\cite{pei09} to a relative magnitude of 10$^{-19}$ under the influence of the 
black-body radiation at room temperature.
In order to avoid the influence of an external magnetic field via the linear 
Zeeman effect, an electronic state has to be chosen for the laser cooling and 
frequency detection, where the total angular momentum F of the atoms is an integer 
number. In this case the Zeeman component $m_F = 0\rightarrow 0$ can be used, 
which shows only a weak contribution of the quadratic Zeeman effect at low 
magnetic fields, typically ranging below 1 kHz at 0.1~mT.
If we can assume an efficient shielding of external magnetic fields (e.g. by 
multiple layers of $\mu$ metal) to below 1~nT, the remaining influence of external 
magnetic fields in $^{229}$Th can be estimated to a relative contribution of 
only 10$^{-18}$.\\
Thus $^{229m}$Th proves to be extraordinary insensitive to external perturbing 
effects in first order for magnetic fields and in second order for electric fields.
This qualifies the isomeric nuclear transition in $^{229}$Th, where the frequency is 
predominantly determined by the strong interaction, as a candidate for a frequency
standard, extremely insensitive against frequency shifts. Therefore, a system could be
realized that is complementary to present atomic frequency standards, which are based
either completely or partly on the electromagnetic interaction with contributions by
the strong force. \\
Additionally, a comparison of the temporal behaviour of the transition frequency of 
$^{229}$Th with these established frequency standards could enable a new opportunity
for a lab-based search for temporal variations of the coupling constants of the
electromagnetic and strong interactions, as it is predicted in some Grand Unified 
Theories. This aspect has been extensively discussed in the literature in recent 
years, including speculations on potential enhancement mechanisms for the 
sensitivity to such temporal variations in the case of 
$^{229}$Th~\cite{fla06,ber09,lit09,fla09,fla09a}.\\
In view of the intriguing potential of the outlined unique properties, already
in the past numerous, however unsuccessful, attempts have been conducted to
directly identify the first excited state of $^{229}$Th.
An obvious reason for this failure follows from the result of a spectroscopic
measurement performed by the Livermore group~\cite{bec07}, which resulted in a 
more than a factor of 2 modified value of the ground-state transition energy
in $^{229m}$Th of 7.6(5) eV, corresponding to a wavelength of 163(11)~nm.
Therefore this transition falls into the region of deep UV, where it could not be
detected by the previous experiments. \\
Certainly the ultimate goal of all studies in $^{229m}$Th is a direct
optical control of the isomeric nuclear transition via laser excitation.
However, this implies an improved knowledge of the transition wavelength by
at least an order of magnitude.\\
In view of the promising properties of $^{229m}$Th, it is not surprising that this
nuclear transition is studied by many groups world-wide.
Different projects are presently pursued, basically following two main experimental
concepts, where the first one builds upon doping host crystals with $^{229}$Th 
and subsequent photo-excitation either directly or via electronic bridge processes. 
The other concept draws on investigating thorium recoil ions from $^{233}$U $\alpha$ 
decay, exploiting the 2$\%$ decay branch to the isomeric state and
targeting an excitation of the isomeric state in an ion trap 
(e.g. linear Paul trap). \\
        At the Lawrence Livermore National Laboratory (USA), where the revised 
        transition
        energy of $^{229m}$Th has been determined, further experiments are ongoing 
        aiming at a measurement of the so far only theoretically estimated 
        isomeric half life of $^{229m}$Th by searching for conversion electrons 
        or photons from conversion decays~\cite{swa10}. Via a population of 
        $^{229m}$Th from the $\alpha$ decay of $^{233}$U the recoil ions are 
        caught and brought between a detector setup of MCP and UV 
        photomultiplier. A determination of the so far unknown half life of
        ${229m}$Th would result in an improved value of the natural line width 
        of the ground-state transition from the isomeric excited state. \\
        The project pursued at the University of California (Los Angeles, USA) 
        draws on the high photon flux of the Berkeley Advanced Light 
        Source (ALS), where based on a 1.9 GeV electron synchrotron 
        10$^{16}$ ph/sec are available at a bandwidth of 0.175 eV. 
        In Ref.~\cite{rel10}, a new 
        approach for a direct identification of the excitation energy of the 
        low-lying isomer in $^{229}$Th was proposed. $^{229}$Th nuclei, implanted
        in a host crystal, should be photo-excited into the isomeric first 
        excited state in the ALS, with subsequent investigation of 
        the induced fluorescence. In order to determine the transition frequency, 
        the photon energy should be scanned in steps of 0.5 eV, thus an 
        envisaged accuracy of the
        transition wave length of about 0.1~nm is targeted. \\
        At Georgia Tech University (USA),
        the development of the laser cooling of highly charged Th$^{3+}$ is 
        in the research focus. For the long-lived $^{232}$Th laser cooling could 
        already be demonstrated~\cite{cam09} and recently a proposal for a
        virtual clock transition composed of stretched states within the 5F$_{5/2}$ 
        electronic ground level of both nuclear ground and isomeric manifolds 
        was proposed and shown to offer unprecedented systematic shift suppression, 
        allowing for clock performance with a total fractional inaccuracy 
        approaching 10$^{-19}$~\cite{cam12}. \\
        At the Physikalisch-Technische Bundesanstalt in Braunschweig 
        (Germany), after a basic study to outline the potential of 
        $^{229m}$Th$^{(3+)}$ qualifying as an ultra-stable nuclear frequency 
        reference~\cite{pei09}, now
        a proposal is pursued towards a novel experimental scheme to excite the 
        nuclear transition $^{229g}$Th - $^{229m}$Th in Th$^+$ ions for a 
        determination of the transition frequency. Based on theoretical 
        calculations~\cite{por10a}, the recently proposed scheme~\cite{por10} 
        employs a bridge process, driven by two incoming laser photons, whose 
        summed frequency resonantly corresponds to the transition frequency. \\
        Also recently a new project started at the 
        TU Vienna (Austria)~\cite{kaz12}. Here $^{229}$Th ions will be implanted 
        into a 
        UV-transparent crystal (e.g. CaF$_2$, LiCaAF$_6$), finally aiming at the
        realization of an optical, solid-state-based nuclear clock. In this 
        concept, the complex vacuum systems of present atomic clocks could be
        replaced by a single crystal, doped with $^{229}$Th atoms at room 
        temperature. \\
        A collaborative project is pursued at the IGISOL facility in Jyv\"askyl\"a
        (Finland) together with the University of Mainz (Germany), 
        aiming at the identification of the 7.6 eV isomer via a measurement of its 
        hyperfine structure~\cite{rae11}. 
        This project is closest to our experimental approach, since here 
        also a $^{233}$U recoil buffer-gas cell is applied~\cite{son12} (see also
        Sect.~\ref{subsec:gas-cell}. 
        Preparatory work has been performed using Resonance Ionization
        Spectroscopy (RIS) to measure isotope shifts of $^{228,229,230}$Th and a 
        template of the ground-state hyperfine structure of $^{229}$Th has been 
        established for a 261.24 nm UV transition.  \\
Moreover, the outlined extensive experimental quest for the low-energy 
$^{229}$Th isomer has also been assisted and guided by comprehensive theoretical 
efforts both within atomic physics and nuclear theory, 
see e.g. Ref.~\cite{lit09,kaz12,tka00,tka03,tka11,kar92,kar05,kar07}.

\section{Experimental approach}
\label{sec:approach}

The experimental approach pursued by our group towards an unambiguous and
largely from potential background contributions separated identification of 
the UV deexcitation of the isomeric excited state in $^{229}$Th is based
on a spatial decoupling of population and deexcitation of the low-lying
first excited state. 
This approach is enabled by the availability of a highly efficient buffer-gas 
stopping cell
that has been developed in Garching. Here the population of the isomeric first
excited state of $^{229}$Th can be achieved via $\alpha$ decay of $^{233}$U.
In order to achieve this, an experimental setup has been realized,
which has been characterized in first preparatory measurements and is being
continuously improved. 
This provides an excellent starting point for a targeted search for the isomeric
ground-state transition in $^{229}$Th and the subsequent development of an optical 
excitation scheme.\\
In the following sections, the various steps of the experimental approach
will be reviewed.

\subsection{Stopping and extraction via a buffer-gas cell}
\label{subsec:gas-cell}

\begin{figure}[htb]
    \centerline{\includegraphics[width=0.5\textwidth,angle=-90]{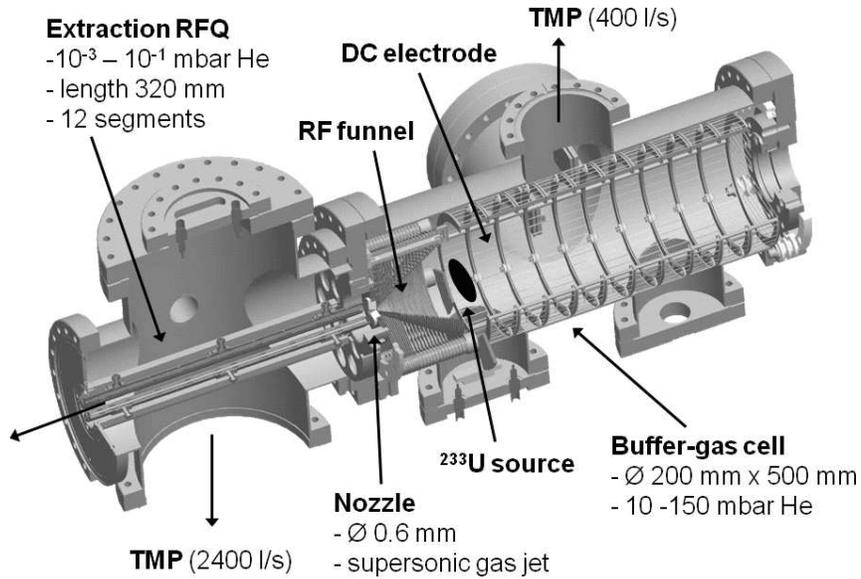}}
        \caption[] {Schematical setup of the buffer-gas stopping cell at 
                    the Garching Maier-Leibnitz-Laboratory (MLL-IonCatcher).}
        \label{fig:mll-ioncatcher}
\end{figure}

Fig.~\ref{fig:mll-ioncatcher} shows the setup of the 'MLL IonCatcher', which is
operated at the Maier-Leibnitz Laboratory in Garching~\cite{NEU06a}. 
It consists of a system of two vacuum chambers, strictly built according to UHV 
standards. The first chamber, the buffer-gas stopping cell, serves to stop energetic 
ions in an atmosphere of ultra-pure helium gas (pressure ca. 25-50 mbar), before 
guiding them via radio-frequency and DC fields towards a nozzle exit, where they 
are dragged out of the chamber via a supersonic gas jet into the subsequent 
extraction chamber. Hereby, the DC guiding field in the gas cell provides a 
controlled acceleration of the stopped ions towards the nozzle. In order to guide 
also ions from regions away from the gas-cell axis towards the nozzle throat, 
a radiofrequency funnel is applied. This is a system of 50 ring electrodes, with 
diameters that conically continuously reduce towards the nozzle. The electrodes are 
electrically isolated with respect to each other and are connected to an RF phase 
alternating by 180$^o$ from one electrode to the next. In addition to the RF 
amplitude, they are provided with a DC gradient in the direction of the nozzle. 
This results in an accelerating force onto the ions towards the nozzle, while the 
radiofrequency field exerts a repelling force of the electrodes onto the ions. 
In total this leads to a funnel-like guidance of all ions to the nozzle exit, where 
the gas flow of the supersonic jet that forms in the nozzle drags the ions off the 
field lines, transporting them into the extraction chamber that houses a 
radiofrequency quadrupole (RFQ) acting as ion guide and phase-space cooler. 
In order to minimize ion losses from recombination or molecule formation inside 
the gas cell, both chambers have been built strictly according to UHV standards, 
i.e. avoiding any organic materials inside the cell by using only stainless steel 
and ceramic components, providing baking capability (up to 180$^o$) and, after 
evacuation to $\le 10^{-11}$~mbar, are operated with catalytically purified 
(gas purity ca. 1 ppb) ultra-clean He 6.0, fed in via electropolished tubes and 
further purified by a cryotrap in front of the cell entrance. \\
The described setup of the buffer-gas cell with its RF- and DC extraction 
electrode systems is shown Fig.~\ref{fig:mll-ioncatcher} together with the 
subsequent extraction chamber and its radiofrequency quadrupole. This RFQ serves 
to transport the ions behind the gas cell, to separate them from the carrier gas 
behind the extraction nozzle and particularly it acts as phase-space cooler for 
the ions behind the gas cell.

High-resolution mass measurements behind the extraction RFQ have proven 
quantitative removal of water and oxygen contaminants, a crucial prerequisite for 
an efficient ion extraction. This buffer gas cell ('MLL-IonCatcher',~\cite{NEU06a}) 
is an optimized second generation of the buffer gas cell built by the LMU group
 for the SHIPTRAP project at GSI~\cite{NEU06}. The SHIPTRAP gas cell is 
routinely very 
successfully used for measurements of high-precision nuclear mass measurements of 
heavy transactinide isotopes at the Penning trap system SHIPTRAP. 
From the experience at SHIPTRAP we know that the dominant fraction of ions leave the 
gas cell in a doubly-charged state due to the low probability of charge-exchange 
reactions in the ultra-pure environment.\\ 
This device enables us to choose a novel approach in the investigation of 
$^{229m}$Th, allowing for an efficient suppression of background contributions 
during the search for the UV fluorescence of the ground-state transition 
of $^{229m}$Th.

\subsubsection{Population and Extraction of $^{229m}$Th using 
                     the gas cell}

\begin{figure}[htb]
   \centerline{\includegraphics[width=0.95\textwidth]{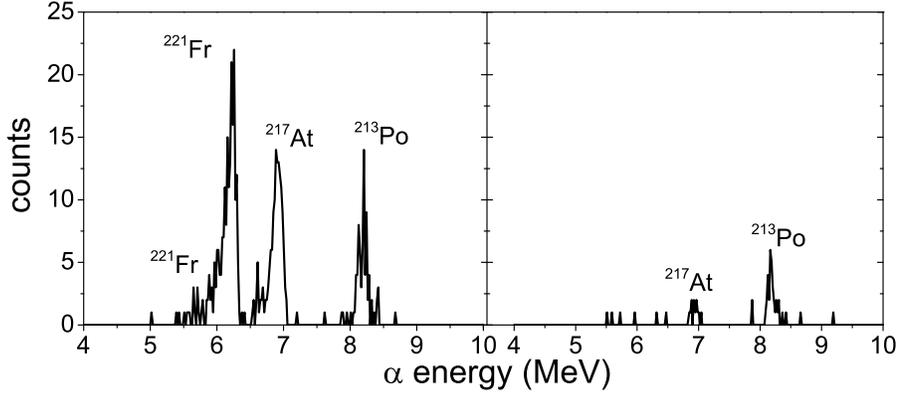}}
        \caption[] {$\alpha$-energy spectrum of daughter nuclei from the decay of
                    $^{233}$U after extraction from the gas cell and collection
                    on a needle tip (left panel).
                    The right panel shows a reference measurement (same measuring 
                    time), where the extracted ions were collected onto a plate located 
                    out of sight from the silicon detector. In this case the
                    direct $\alpha$-decay line from $^{221}$Fr disappears and only
                    and only a weak component from $^{213}$Po remains, originating 
                    from the decay of $^{217}$At recoil ions, implanted during the decay 
                    of $^{221}$Fr at sites visible from the detector position.}
        \label{fig:e-alpha-rfq-needle-si-plate}
\end{figure}

While presently a direct population of the thorium isomer via laser excitation
still fails due to the extremely narrow line width of the transition together
with an rather unprecise knowledge of the transition energy, instead
the $\alpha$ decay of $^{233}$U to $^{229}$Th can be used for a population of
the isomer. This is provided by a decay branch with an intensity of about
2$\%$, leading from $^{233}$U to the first excited state of $^{229}$Th.  \\
We placed a radioactive $^{233}$U source (evaporated onto a steel backing) 
inside the above described 'MLL-IonCatcher' 
buffer-gas stopping cell (presently using an effective source activity of 5 kBq), 
such that the $^{229m}$Th $\alpha$ recoil nuclei produced 
during the $\alpha$ decay will be stopped in about 40 mbar helium and then will 
be guided by the RF and DC fields as described before to the nozzle exit of the 
gas cell. Here they are transferred by the gas flow into the vacuum regime of the 
subsequent RFQ phase-space cooler and ion transport channel. The extraction from 
the gas cell can be performed within 1-2 ms. \\ 
This experimental arrangement assures that all background processes like
conversion or prompt excitations happen inside the gas cell, while only 
$\alpha$ recoil nuclei will be extracted, thus eliminating
the contamination by prompt background processes.\\
From the decay chain of $^{233}$U it turns out that besides $^{229m}$Th also 
other short-lived $\alpha$ recoil nuclei
can be expected behind the gas cell ($^{221}$Fr, $^{217}$At, $^{213}$Po). 
For the identification of extracted $\alpha$ recoil nuclei from the gas cell
and to demonstrate the feasibility of collecting the recoil ions after the
extraction RFQ, in a first exploratory setup a steel needle tip set on an 
appropriate attractive potential to capture the extracted (positive) ions was 
mounted behind the RFQ, while a silicon particle detector was positioned sideways 
without direct sight to the RFQ exit. The resulting $\alpha$ energy spectrum is 
shown in Fig.~\ref{fig:e-alpha-rfq-needle-si-plate}.

We find in the left panel the $\alpha$ spectrum of $^{233}$U daughter nuclei. 
As a proof that it indeed originates from $\alpha$ decays on the needle 
tip and not from otherwise implanted recoil nuclei, a reference measurement was
performed, where the collection potential was chosen such that ions were
accumulated on a plate that in the previous case seved as electrostatic reflector
to achieve a soft collection of ions on the needle tip. Now the silicon 
detector was mounted without direct sight to the collection surface.
This leads to the spectrum shown in the right panel, 
where the previously dominant line from $^{221}$Fr has disappeared and only 
a weak component from $^{213}$Po remains, originating from the decay
of $^{217}$At ions, implanted during the decay of $^{221}$Fr at sites visible 
from the detector position.\\
This proves in principle the feasibility of the described production scheme 
for $^{229m}$Th, while still calling for an optimization of the collection
 efficiency and a purification of the extracted ions from decay products other 
than $^{229}$Th. \\
In a first step the extraction efficiency out of the gas cell was determined
using a $^{223}$Ra $\alpha$ recoil source. From the identification of the
$\alpha$ decay of the $^{215}$Po decay products, an extraction efficiency 
of 48$\%$ could be determined for this ion species~\cite{thi07}.
However, we do not claim this efficiency also to be valid for the extraction
of $^{229}$Th in view of experimental findings at the IGISOL facility in 
Jyv\"askyl\"a~\cite{son12}.
There a rather high reduction factor of 10 was observed between the efficiency
measured for $^{221}$Fr$^+$ and $^{229}$Th$^+$, which was assigned to reactive
impurities in the buffer gas like oxygen or water. Since cleanliness in the
gas cell as well as for the used He gas was a design criterion from the start
and could already be demonstrated in sensitive mass spectrometry, exceeding the
provisions included in the IGISOL setup in several aspects (efficient RF funnel
structure, baking capability to $\sim$ 180$^o$, electropolished tubing
with VCR fittings, catalytic purification of highest-grade He 6.0),
we are confident that it is fair to assume a reduced loss factor of 5 instead 10 
for the MLL-IonCatcher, thus ending with about 10$\%$ extraction of $^{229}$Th.
However, as shown later, even without this factor of 2 higher extraction
efficiency, enough detectable photon rate and signal contrast could be achieved.
Moreover, we assume 70$\%$ population of the 2$^+$ charge state, derived from our
experience at the SHIPTRAP gas cell, where Penning-trap mass measurements 
routinely use 2$^+$ ions from the gas cell. Thus our best estimate for the expected
extraction efficiency of $^{229}$Th$^{2+}$ is 7$\%$.\\
While we are presently limited by the total $^{233}$U source activity of 15 kBq,
corresponding to a usable Th recoil intensity of 5.0 kBq and finally ca. 
7 extracted $^{229m}$Th isomers per second, we have in the meanwhile 
obtained a license to handle a total source activity of up to 260 kBq, thus opening 
the perspective to further increase the Th flux by a factor of about 16 (with the 
final limiting factor being the maximum diameter of about 90 mm for the $^{233}$U 
source inside the gas cell). Producing such a large-area source via electro-plating
requires great care to achieve good surface homogeneity in view of the thin
$^{233}$U layer to be realized (ca. 13 nm). Moreover, one has to avoid any chemical
contaminations from the source production process on the source surface that could
be transferred as gas impurities into the ultra-pure He buffer gas via surface 
sputtering during the $\alpha$ decay process. Here one has to take into account 
already the unavoidable oxide contamination introduced by using uranium oxide 
instead of the more favourable, however highly oxidizing metallic uranium.

\subsection{RF mass filter for exclusive extraction of $^{229}$Th
                    $\alpha$ recoil nuclei}
\label{subsec:qms-new}

\begin{figure}[htb]
    \centerline{\includegraphics[width=0.9\textwidth,angle=-0]
               {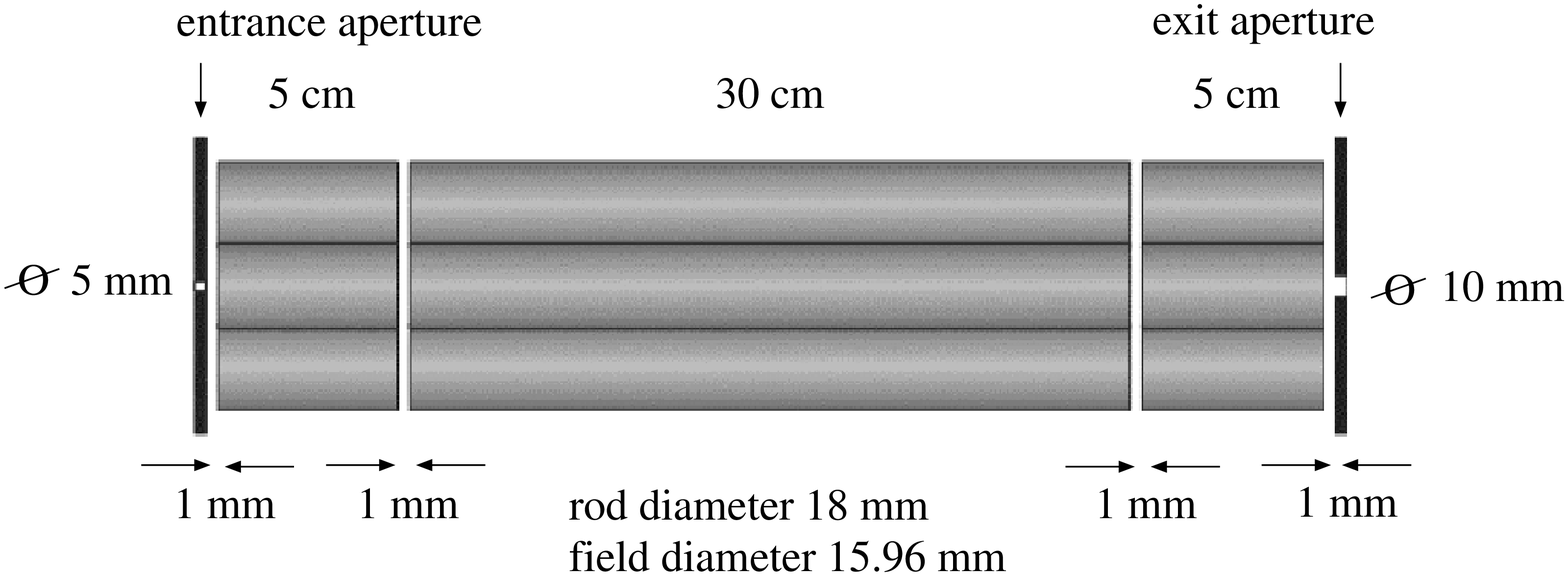}}
    \vspace{0.3cm}
    \centerline{\includegraphics[width=0.6\textwidth]{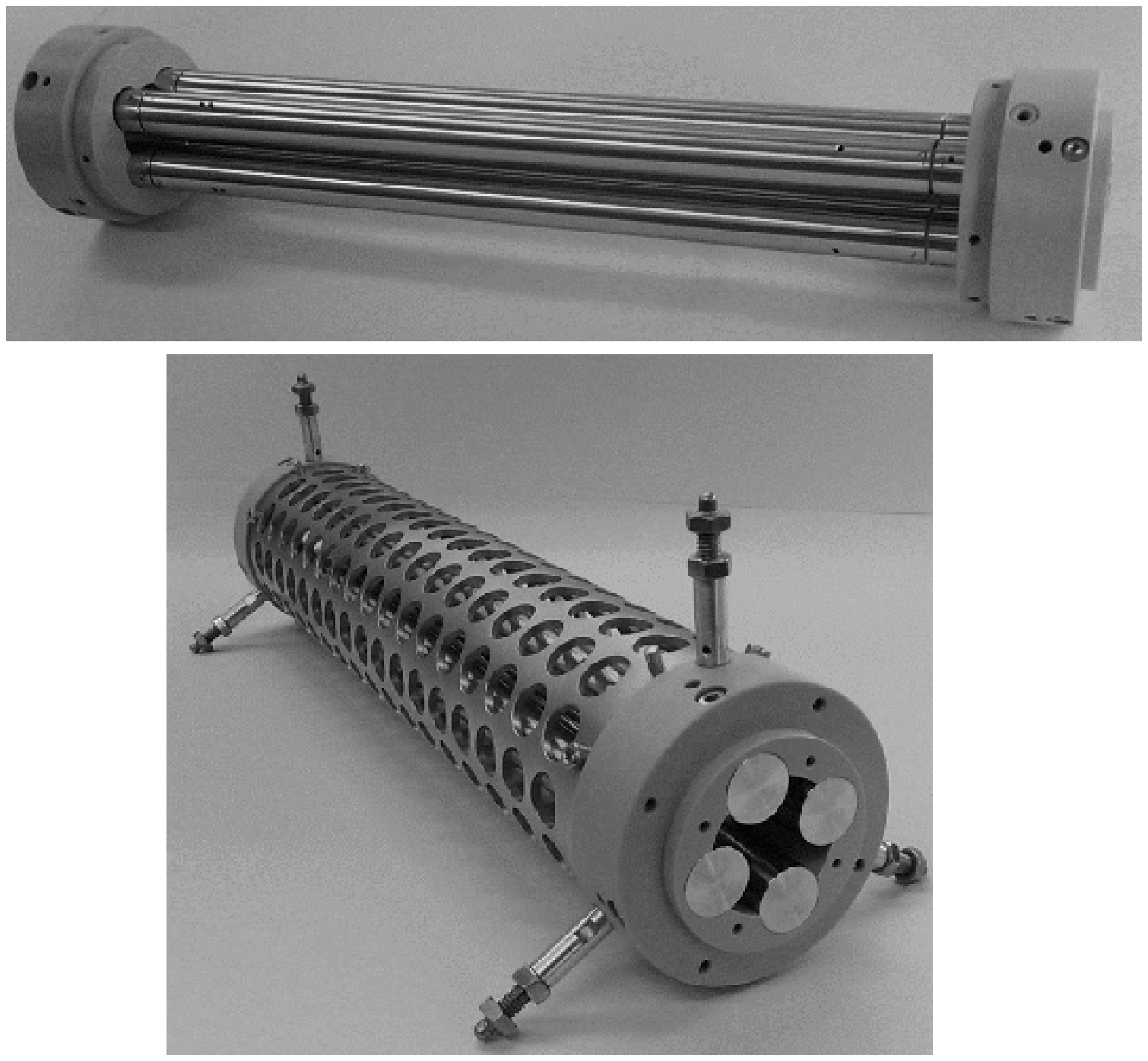}}
       \caption[] {Quadrupole mass filter for an exclusive 
                   extraction of $^{229(m)}$Th from the gas cell after 
                   $\alpha$ decay of a $^{233}$U source. Upper and middle
                   panel: sketch and photograph of the rod system (including
                   Brubaker lenses at entrance and exit sides), bottom part:
                   photograph of the full system including shielding tube and
                   positioning posts.}
        \label{fig:qms-photo}
\end{figure}

From the $\alpha$ energy spectrum (Fig.~\ref{fig:e-alpha-rfq-needle-si-plate}) 
measured behind the gas cell it is obvious, that not only the direct $\alpha$  
daughter nucleus $^{229}$Th, but also the further decay products $^{225}$Ra and 
$^{221}$Fr can be extracted from the gas cell, potentially also acting as sources 
of subsequent UV fluorescence. 
Only with a full control over the recoil ions, as realized in our experimental
approach of the ion transport through an RF ion guide to the detection section,
an additional selection stage can be implemented, able to guarantee the 
requested unambiguity of the investigated ion species.
In order to achieve this goal, the experimental setup has been extended by a 
quadrupole mass filter (QMS) behind the extraction RFQ. A first version that 
already achieved the required mass resolution of $\Delta m \sim 1$ 
(however at the expense of transmission efficiency) was recently replaced by a 
new device that was constructed following a novel and optimized design developed 
by the Giessen group~\cite{haettner11}, where besides utmost mechanical precision 
(tolerance of rod distances: 10 $\mu$m) Brubaker lenses~\cite{brubaker68} have been 
added at the entrance and exit of the QMS in order to allow for an optimized 
transmission efficiency. Fig.~\ref{fig:qms-photo} shows a sketch and a photograph 
of the QMS rod system (upper and middle part) and the complete device including the 
shielding tube and positioning posts (bottom part).

Hereby, a selective transport of Th ions with mass 229 will be enabled, with
simultaneous efficient suppression of lower masses, particularly 225 (Ra) 
and 221 (Fr). Thus the achievable mass resolution of about $m/\Delta m\sim$ 160
with the new QMS will be sufficient to guarantee exclusive extraction of
$^{229}$Th.\\
A prerequisite for this mass resolution is a precise adjustment of the two 
alternating RF phases at the QMS. An active (LabView-based) control system 
was implemented and successfully commissioned, where the actual RF amplitudes, 
read out via a precise (12-bit) PC-based oscilloscope emulation, where adjusted 
to each other by trimming a remotely tunable capacitor. Thus a relative accuracy 
of the two RF amplitudes of $\le$ 1$\cdot 10^{-3}$ could be reached~\cite{kalb12}.\\
In order to characterize and optimize the mass filter, an alkali test ion source
was integrated into the gas cell, able to provide $^{39,41}$K, $^{85,87}$Rb and 
$^{133}$Cs ions with sufficiently high rate for an optimization of the gas cell
and QMS operational parameters.
For these measurements an MCP detector for an individual registration
of extracted ions is used instead of the Si detector used for the detection 
of $\alpha$ decays.
Already with the predecessor version of the new QMS, a clear separation of the 
neighbouring mass peaks from $^{85,87}$Rb could be achieved, while now
the new device, will allow
for a mass resolving power $m/\Delta m\sim$ 160 (measured at 10$\%$ of the mass 
peak height) together with a transmission efficiency of about
80-90$\%$ as demonstrated in Ref.~\cite{haettner11}. Thus the suppression 
of the accompanying $\alpha$ recoil nuclei $^{225}$Ra and $^{221}$Fr, which is 
mandatory for a selective extraction of $^{229(m)}$Th from the gas cell, can safely 
be realized. 

\subsection{Setup of ion collection system}
 \label{subsec:collection}

\begin{figure}[htb]
   \centerline{\includegraphics[width=0.90\textwidth,angle=0]{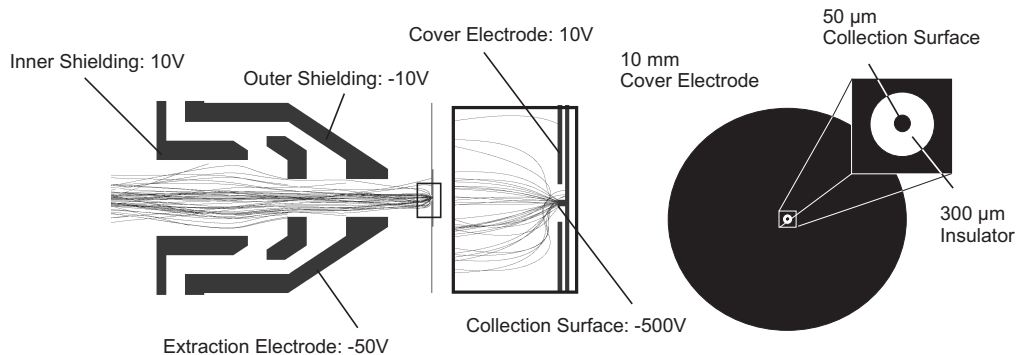}}
   \caption[] {Schematical drawing of the nozzle-like electrode setup together with
               simulated ion trajectories
               (left) to focus ions entering from the QMS for efficient collection 
               onto a small collection surface (right). In the middle part an enlarged 
               view (corresponding to the region inside the marked rectangle) is
               displayed to highlight the focusing and collection properties.}
   \label{fig:collection-nozzle}
\end{figure}

Ion trajectory simulations using the SIMION code package have been performed to
design an optimized ion collection system behind the QMS exit by studying
a wide variety of geometrical arrangements. During the simulations,
the ions were followed over their full trajectory, starting from the
$^{233}$U source inside the gas cell, over the extraction RFQ and the quadrupole 
mass filter up to the final collection stage. Thus particularly the divergence
of the ions exiting the QMS, which critically determines the achievable collection
efficiency, could be realistically included in the optimization process.
Design goals were (i) to achieve the highest possible ion collection efficiency
and (ii) to realize the smallest possible collection surface in order to
provide an almost point-like source for the targeted UV emission. Such a source 
will allow for a subsequent UV optics that minimizes the transmitted source image
size onto the MCP detector in order to optimize the signal-to-background ratio.\\
Here the initial solution of attracting the ions via an electrical mirror
onto a small (diameter 40 $\mu$m) steel needle tip positioned perpendicular 
to the extraction axis turned out to be inferior to another solution, where the 
ions are collected on a metallic surface (diameter 50 $\mu$m) directly facing the 
extraction axis. Moreover, in order to achieve a rather high ion collection, an 
attractive collection potential of -500~V needs to be applied at the collection 
surface. The simulated collection efficiency amounts to 40$\%$, achieved by using a 
nozzle-like focusing electrode system behind the QMS exit as shown in 
Fig.~\ref{fig:collection-nozzle} (left part). The middle part of 
Fig.~\ref{fig:collection-nozzle} shows an enlarged view of the rectangle marked
behind the nozzle exit, indicating the focusing and ion collection capabilities
of the collection surface held at -500~V.

The right part displays a front view of the collection surface (50~$\mu$m in 
the center, detailed view given in the inset), 
which is surrounded by an insulating rim (diameter 300 $\mu$m) and
an outer cover electrode (diameter 10 mm). This structure has been manufactured
by laser-drilling and etching from a polyimide laminate PCB, Cu-coated on both
 sides.

\subsection{Provisions against quenching processes}
  \label{subsec:quenching}

When discussing the de-excitation of the excited isomeric $^{229m}$Th state,
we have to take into account the possibility of a non-radiative deexcitation,
i.e. a quenching process originating from a surface interaction between the
collected (excited) Th ions ('donor') and the collection surface
('acceptor')~\cite{tka00}.
Several mechanisms of this energy transfer are usually described in the context
of fluorescence chemistry. In our case the F\"orster resonance energy transfer
(FRET)~\cite{foe48} will be the most important one. The FRET mechanism is based
on a dipole-dipole interaction between the donor and the acceptor. Due to this
interaction a virtual photon, which violates energy conservation, can be
interchanged over distances of up to 10 nm. Within a QED description, the
possibility of quenching is connected to the emission and absorption spectral
overlap between donor and acceptor. In case that no such spectral overlap exists,
no quenching can occur~\cite{hau69}, meaning that in our experiment quenching
can be ruled out in case the collection of the isomers takes place on a surface
(of thickness larger than 10 nm) with no spectral absorption in the vacuum-UV
region around 160~nm. This can be achieved by coating the collection surface,
e.g., with MgF$_2$, which is a standard material used for protective coatings
of Al mirrors. Thus we propose an MgF$_2$-coated metallic surface as ion
collection electrode, where the thin dielectric
layer will not affect the collection properties of the electrode operated at
$\sim$-500~V, e.g., by surface charging effects~\cite{bud68} due to a small current
in the dielectric material, large enough to a direct discharge.

In order to obtain a more quantitative assessment of radiationless deexcitation
processes, we propose to measure the effect of quenching in a separate
experiment, e.g. by using atomic species with UV transitions that are easier to
access compared to $^{229}$Th. This way the amount of quenching could be
quantified and the above proposed method to prevent radiationless deexcitation
could be tested.

\subsection{Optimizing the UV focusing optics}
\label{subsec:focusing}

\begin{figure}[htb]
   \centerline{\includegraphics[width=0.9\textwidth]{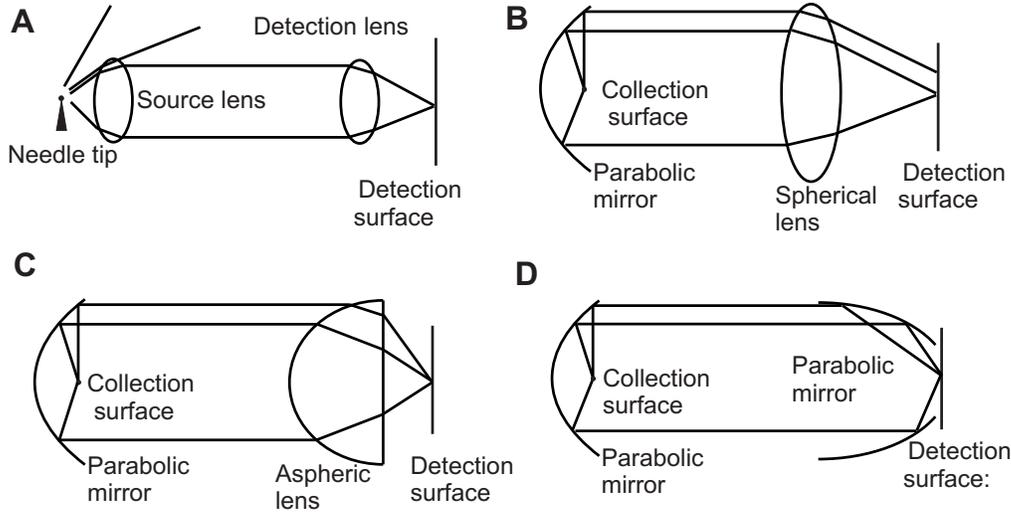}}
        \caption[] {Schematical drawings of the four UV optical arrangements
                    studied to focus the $^{229m}$Th UV fluorescence onto an
                    MCP detector.}
        \label{fig:optical-systems}
\end{figure}

Extensive simulations using an optical ray-tracing code, specifically
developed for this purpose, were performed to benchmark different options for
the UV optical system.\\
Optimization criteria were (i) a maximized photon collection efficiency and
(ii) a small image magnification to achieve the
best possible signal contrast ratio, the latter imposing strict requirements on
the focal lengths of the optical elements, leading to contradictions with large
numerical apertures favoured for (i) when using lens-based systems. Moreover,
spherical aberrations also limit the performance of setups based on spherical 
optical elements. This 
naturally leads to consider setups based on parabolic mirrors, which are 
favourable compared to aspherical elements due to their larger acceptance and 
broader wavelength dynamics.\\
Possible arrangements to focus the UV radiation that have been comparatively
studied are (A) a system of two convex lenses,
(B) a system consisting of a parabolic mirror and a (spherical) convex lens,
(C) a system consisting of a parabolic mirror and an aspherical lens and (D)
a system of 2 parabolic mirrors (as schematically shown in
Fig.~\ref{fig:optical-systems}), each representing the best what can be
achieved within the considered type of optical arrangement. \\
It should be noted that the double-lens system in contrast to the mirror-based
scenarios would still require the (highly inefficient) ion collection on a
needle tip.

Numerical results obtained for various properties of these optical systems are
listed in Tab.~\ref{tab:optical-scenarios}. They represent the result of extensive
calculations optimizing the optical system in the context of the full experimental
setup in all operational parameters.

\begin{table}[htb]
 \caption[]{Properties of the four alternatively studied optical scenarios
            to focus the UV fluorescence radiation: (1) percentage of photons
            reaching the MCP, starting from source emission as 100$\%$; (2)
            percentage of photons focused into image spot, taking all photons
            reaching the MCP as 100$\%$; (3) percentage of photons focused
            into image spot, starting from source emission as 100$\%$;
            (4) same as (3), but including lens transmission efficiencies (0.85)
            and mirror reflectivities (0.7); (5) absolute photon counting rate
            on MCP, starting from 2.24 isomer decays/sec. from collection surface
            (expected for case B, C and D), including MCP efficiency
            of 0.16 at 160 nm; (6) image spotsize obtained from numerical
            calculations, including source diameter and exact optical shapes;
            (7) image magnification as derived from source dimensions; (8)
            count rate divided by the expected image area; 
            (9) image intensity divided by
            expected dark count rate of 0.05/s mm$^2$. The values listed for
            scenario C are given for a wavelength of 163 nm and degrade when
            deviating from this design value (in contrast to the performance
            of the double-mirror scenario D) }
    \begin{tabular}{l|ll|ll}
          &\multicolumn{2}{|c|}{A: Double (spherical) lens} &
           \multicolumn{2}{|c}{B: Mirror / (spherical) lens}  \\
                           & 10$\%$ & FWHM & 10$\%$ & FWHM \\
    \hline
 geometric acceptance [$\%$] (1)  & 2.4  &  2.4 & 40.5 & 40.5  \\
 geom. collection efficiency [$\%$] (2)& 40.4 & 10.1 & 79.0 & 25.5 \\
 optical efficiency [$\%$] (3)    & 0.99 &  0.25& 32.0 & 10.3  \\
 photon collect. efficiency [$\%$] (4)& 0.72 &  0.18& 19.0 &  6.1 \\
 counting rate [s$^{-1}$] (5)  &0.0026& 0.00064&0.068& 0.022  \\
 \hline
 image spotsize [mm] (6)  & 0.17 & 0.045& 0.51 & 0.21  \\
 effective image magnification (7)   &  3.5 &  0.9 & 10.2 & 4.3   \\
 image intensity [s$^{-1}$mm$^{-2}$] (8)&0.11 & 0.40 & 0.34 & 0.61 \\
 signal contrast (9)         & 2.2 &  8.0 & 6.8  & 12.2  \\
\hline
\hline
    \end{tabular}
    \begin{tabular}{l|ll|ll}
                    &  \multicolumn{2}{|c|}{C: Mirror/aspherical lens} &
                            \multicolumn{2}{|c}{D: Double mirror}\\
                    & 10$\%$ & FWHM & 10$\%$ & FWHM \\
    \hline
 geometric acceptance [$\%$](1)  & 41.4 &41.4 & 41.3 & 41.3 \\
 geom. collection efficiency [$\%$](2) & 78.8 & 19.3 & 86.7 & 60.3 \\
 optical efficiency [$\%$](3)    & 32.6 & 8.0 & 35.8 & 24.9 \\
 photon collect. efficiency [$\%$] (4) & 19.4 & 4.8 & 17.5 & 12.2\\
 counting rate [s$^{-1}$] (5)  & 0.069 & 0.017 & 0.063& 0.044 \\
 \hline
 image spotsize [mm] (6)   & 0.067& 0.021 & 0.066& 0.044 \\
 effective image magnification (7)   & 1.33 & 0.42 & 1.33 & 0.87  \\
 image intensity [s$^{-1}$mm$^{-2}$] (8) & 19.9 & 50.7 & 18.1 & 29.3 \\
 signal contrast (9)         & 398 & 1015 & 362   & 586 \\
\hline
    \end{tabular}
  \label{tab:optical-scenarios}
\end{table}

The image spot size is indicated here as the diameter of the photon distribution 
on the MCP detector given as FWHM value or at 10$\%$ of the peak maximum.
From all scenarios studied, our favoured solution (scenario 'D' in
Fig.~\ref{fig:optical-systems}) is an arrangement, where the
collection surface is positioned in the focus of an annular parabolic mirror
(dielectric coated aluminum mirror, outside diameter 40 mm, leaving a
central hole of 12 mm diameter), which
parallelizes the emitted UV light and transmits it onto a second parabolic mirror.
This deep parabolic mirror is shaped via a central hole such that its retracted
focal point lies
in the plane of the MCP detector positioned directly behind the mirror.
With an optical efficiency of 35.8$\%$ and typical reflectivities at 160~nm of
70$\%$ for each of the two mirrors, a total UV photon collection efficiency of 
17.5$\%$ can be achieved.\\
This arrangement allows for a small image size of ca. 70 $\mu$m diameter,
concentrating the UV flux onto only about 50 MCP pixels. Thus a very high signal 
contrast (see line (8) in Tab.~\ref{tab:optical-scenarios} comparing to the 
MCP dark count rate) can be achieved. We also studied a scenario based on the 
first parabolic mirror and an aspherical lens replacing the second mirror, 
indicated as scenario 'C'. At first glance it appears as if this arrangement were 
even superior to the double-mirror setup discussed before (mostly owing to the
difference between the assumed lens transmission of 85$\%$ compared to the mirror
reflectivity of 70$\%$, which, however, may be over-estimated for the actual
thickness of 20 mm needed for the aspherical lens in our case). In fact, both
scenarios 'C' and 'D' draw on the correction of the spherical aberration to
reach their impressive performance. However, it should be emphasized that the
values listed in Tab.~\ref{tab:optical-scenarios} for setup 'C' only hold for
the design wavelength of 163 nm, while significant degradation is found when
deviating from this wavelength. As an example, considering a wavelength deviation 
of $\pm$10 nm, i.e. calculating setup 'C' for 153 nm and 173 nm, respectively,
brings down the signal contrast from 398 (at 10$\%$ maximum height) to
154 (153 nm) or 176 (173 nm), ending up about a factor of 2 inferior to the
double-mirror setup. Moreover, one has to keep in mind that for an optimum
performance of setup 'C' the focal position of the aspherical lens has to be
defined with a precision of about 10 $\mu$m, which may turn out to be a time
consuming effort in view of the about 2 mm shift of the focal length when
varying the wavelength between, e.g., 153 nm and 173 nm, respectively. Assuming
a step width of 50 $\mu$m, a maximum of 40 measurements will be required to
search for the lens focus. Particularly when taking into account potential
radiationless deexcitation processes and thus reduced detectable UV photon yield, 
the individual measurement time needed for each of these steps may rise up to 
50 hours, rendering the search procedure a lengthy and maybe even prohibitive task.
In view of the comparable count rates that can be achieved in either of the two 
cases (once the correct focal position of the aspherical lens has been found), 
we consider both setups as viable options, where technical feasibility of the 
respective optical elements will finally decide on the realization, nevertheless 
with a clear preference for the double-mirror setup due to its less critical 
wavelength dependence as outlined above. \\
During the commissioning phase, a deuterium lamp (with a UV emission
spectrum around 160~nm) will be positioned behind a 50$\mu$m pinhole
in the collection chamber.
This will allow for a characterization and optimization of the optical
transmission and detection properties.

\subsection{Detection of the $^{229m}$Th UV fluorescence radiation}
  \label{subsec:detection}

In order to detect the UV radiation originating from the deexcitation
of the $^{229m}$Th isomeric state captured on the collection surface described 
before, the above described UV optical system will be placed behind the collection  
surface, allowing to focus a large fraction of the 7.6 eV UV fluorescence light 
onto a CsI-coated, UV-sensitive multi-channel-plate detector (MCP: 75~mm diameter 
in Chevron geometry, channel diameter 10$\mu$m). The electrons 
that are generated in the MCP will then be accelerated onto a phosphor screen, 
where they are converted into visible light, subsequently registered by a highly 
sensitive, low-noise CCD camera.\\
In contrast to a simpler detection scheme using a single photomultiplier, 
the pixel structure of the MCP (ca. 10$^6$ electron amplifier channels per cm$^2$)
offers the opportunity to discard a majority of the dark count rate, provided
an optimized focusing of the UV fluorescence.
Typical dark count rates of MCP detectors of 0.05 counts/(sec. $\cdot$ mm$^2$)
result in a dark count rate of 5$\cdot 10^{-6}$ per pixel and second. When we 
therefore achieve to focus the image of the UV source onto 50 pixel of the 
MCP (corresponding to an image diameter of about 70 $\mu$m, which we expect 
from our simulations as diameter of the image spot), this corresponds to a 
background contribution to the signal strength in the image spot due to the 
MCP detector of 2.5$\cdot 10^{-4}$/sec. This low dark count rate can by far 
not be reached by any low-noise single photomultiplier.

\subsection{Improved determination of $^{229m}$Th ground-state 
                transition}
  \label{subsec:energy}

\begin{figure}[htb]
   \centerline{\includegraphics[width=1.0\textwidth,angle=0]
               {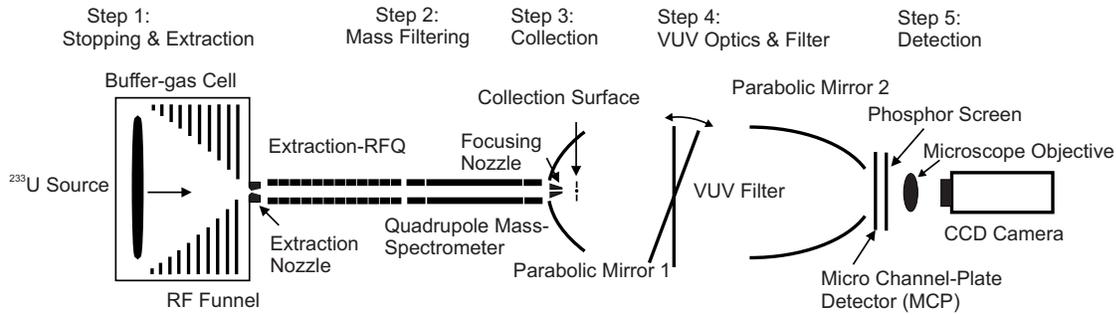}}
        \caption[] {Experimental scheme for the identification of UV 
                    fluorescence from the deexcitation of the isomeric 
                    first excited state of $^{229m}$Th.}
        \label{fig:setup-complete}
\end{figure}

The optical double-mirror setup introduced before has the additional advantage 
that it leaves ample space between the mirrors to install a filter system
that can be used to determine the transition wavelength of $^{229m}$Th.
Since the optical pathway will be parallel between the two mirrors, no effects
from convergent or divergent optical paths through the mirrors have to be  
considered. Here we intend to apply a straightforward 'binary search' technique.
As soon as the fluorescence signal from the $^{229m}$Th decay has been 
unambiguously identified, the next project step will target to significantly 
improve the accuracy of the transition wavelength to better than 1~nm, 
corresponding to an improvement of about one order of magnitude compared to our 
present knowledge.
This can be achieved by inserting special UV absorption filters with
particularly sharp absorption edges into the direction of the 
focused fluorescence radiation, searching for a disappearance of the 
7.6 eV signal. Fig.~\ref{fig:setup-complete} visualizes this concept.

For this purpose, two specially coated VUV filter sets with absorption edges near
160~nm and 170~nm, respectively, with a width of the absorption edge of ca. 1~nm
have already been purchased.
More filters with fine-tuned absorption edges should be acquired depending on the
coarse location of the transition energy determined with the existing filter sets.
Moreover, using these filters, the position of the absorption edge can be varied 
by a variation of the entrance angle as indicated in 
Fig.~\ref{fig:setup-complete}. In order to achieve this the filter will be
positioned on a remotely controlled rotary stage.\\
With this rather straightforward method it will be possible to improve the
present accuracy of the ground-state transition energy of the isomeric first excited state 
in $^{229}$Th by about an order of magnitude, thus providing the prerequisites
for the development of a dedicated laser system.

\section{Concluding efficiency assessment and count 
           rate estimate}
 \label{sec:efficiency}

Having quantitatively discussed all components of the experimental setup, we can 
finally present the full picture, outlining the various contributions to the
final detection efficiency and specifying the expected count rate and signal 
contrast.\\
{\bf $^{233}$U $\alpha$ source}: presently we are operating a $^{233}$U source with
   an effective Th activity of 5.0 kBq. In the meanwhile, we have successfully
   applied 
   for an increased handling license up to 260 kB total activity. In view of the
   2$\pi$ acceptance for recoil emission, taking into account the recoil stopping
   range of Th in U (ca. 10 nm), the actual chemical stoechiometry of
   the source material (U$_3$O$_8$) and considering the geometrical constraints 
   inside
   the gas cell (limiting to a source diameter of 90 mm), a further increase of 
   the
   effective (angle-integrated) Th flux by a factor of 16 up to 80 kBq can be 
   achieved. Together with the
   2$\%$ decay branch to the first excited isomeric state, presently 100 
   (future: 1600) 
   $^{229m}$Th ions per second enter the buffer gas stopping volume. 
   It should be emphasized that the intended increase of the source area will 
   not affect
   the extraction efficiency due to the guiding and focusing properties of the 
   RF funnel. \\
{\bf Buffer-gas cell:} As discussed in Sect.~\ref{subsec:gas-cell}, we expect an 
   extraction efficiency from the buffer-gas cell for $^{229m}$Th$^{2+}$ of 7$\%$.\\
{\bf Mass filter:} The QMS transport efficiency will amount to about 80$\%$.\\
{\bf Ion collection:} An ion collection efficiency of 40$\%$ is expected behind 
     the QMS.\\
{\bf UV fluorescence focusing:} according to the discussion in
   Sect.~\ref{subsec:focusing}, about 17.5$\%$ of the emitted 7.6 eV photons will 
   be focused by the double-mirror arrangement onto the detection device.\\
{\bf MCP detector:} The detection efficiency at 160~nm for a CsI-coated and thus 
     UV-sensitive MCP detector typically amounts to 16$\%$. \\

Finally, this leads to total efficiency of 1.2$\cdot$ 10$^{-5}$, corresponding
to a detectable UV photon rate of 0.06/s for the present 5.0 kBq Th source,
which can be increased to 0.96/s for a source activity of 80 kBq as envisaged
in the near future.
In view of the focal spotsize on the MCP of 70 $\mu$m diameter (area ca. 
0.004 mm$^2$)
and a typical MCP dark current of 0.05 counts/(s$\cdot$mm$^2$), even the present
source strength would allow for a signal:background-ratio of 362:1, which could be
increased further by a stronger source up to ca. 5800:1.\\
It should be stressed that with the present $^{233}$U source and the double-mirror 
setup,
already a superb signal contrast can be achieved, which is expected to be high 
enough 
to detect the $^{229m}$Th de-excitation radiation even if significant non-radiative 
losses via bridge processes or conversion decays in the range of 90$\%$ occur.
With the envisaged increased source activity by a factor of 16, our detection
sensitivity could be increased to even as low as 1$\%$ of the expected 
deexcitation processes. \\

Thus we conclude that we present here an experimental
concept that has been studied and designed quantitatively in all components, 
either via preparatory experimental work or by detailed simulations, ending
in a scenario with high enough sensitivity to achieve a detection of the 
UV fluorescence from the low-lying isomeric first excited state of $^{229}$Th even
in case of unfavourable decay branchings and to target an improvement of the
experimental accuracy of the ground-state transition.

\acknowledgments
  We acknowledge fruitful discussions with T.W. H\"ansch, T. Udem, D. Habs, 
  E. Haettner, P. Hilz and J. Schreiber. 
  This work was supported by the DFG Cluster of Excellence Munich-Centre for 
  Advanced Photonics (MAP).

\end{document}